\documentstyle[aps,pre,epsfig,amssymb]{revtex}

\begin{document}

\draft

\title{Roughening Transition of a Restricted Solid-on-Solid Model \\
       in the Directed Percolation Universality Class}

\author{J. Ricardo G. de Mendon\c{c}a\cite{email}}

\address{Departamento de F\'{\i}sica, Universidade Federal de S\~{a}o Carlos,
13565-905 S\~{a}o Carlos, SP, Brazil}

\date{Physical Review E {\bf 60}, 1329--1336 (1999)}

\maketitle

\begin{abstract}
We carried out a finite-size scaling analysis of the restricted 
solid-on-solid version of a recently introduced growth model that 
exhibits a roughening transition accompanied by spontaneous symmetry 
breaking. The dynamic critical exponent of the model was calculated 
and found to be consistent with the universality class of the directed
percolation process in a symmetry-broken phase with a crossover to 
Kardar-Parisi-Zhang behavior in a rough phase. The order parameter of 
the roughening transition together with the string order parameter was
calculated, and we found that the flat, gapped phase is disordered
with an antiferromagnetic spin-fluid structure of kinks, although 
strongly dominated by the completely flat configuration without kinks. 
A possible interesting extension of the model is mentioned.
\end{abstract}



\section{INTRODUCTION}

It has been realized that nonequilibrium interacting 
particle systems are capable of exhibiting very interesting and unusual 
phenomena in (1+1) dimensions, such as single-defect and boundary induced 
phase transitions \cite{krug,kandel,efgm}. Moreover, it is known that some 
of these phase transitions are accompanied by spontaneous symmetry breaking 
(SSB) \cite{efgm,arndt}, in which some macroscopic observable of the model
behaves in the steady state asymmetrically with respect to what it would be
expected from the microscopic rules governing its dynamics. All this is rather
unusual for equilibrium one-dimensional systems of particles interacting 
through short range forces only, and there have been serious attempts towards
the understanding of these phenomena, in particular that of SSB \cite{bennett}.
It appears that among the many ingredients favoring SSB to take place,
unbounded noise is a key one, whether it comes from the introduction of 
defects, boundary terms or any microscopic rules rendering a system lacking
in detailed balance.

Growth models provide a suitable theoretical framework for the investigation 
of a gamut of model systems \cite{halpin}, and it turns out that they can be
used to investigate the above mentioned  phenomena as well. In a recent 
work \cite{aehm}, a class of growth models addressing both the questions of 
a roughening transition in one dimension and of SSB was introduced. 
Concerning SSB, these authors have inquired about the necessary ingredients 
in a model in order for it to show SSB, and they found it possible to 
have SSB characterized by a nonconserved order parameter in a ring geometry,
as opposed to the situation in a related model showing the same characteristics
but now with the SSB associated with a conserved (i.e.,~slow) order parameter
and in the presence of conspicuous boundary terms \cite{efgm,arndt}. In the
unrestricted version, the models can be related to a directed percolation
process thus sharing its exponents, a fact that was confirmed by Monte Carlo
simulations \cite{aehm}. In the restricted solid-on-solid (RSOS) version no
such relationship exists. It is possible, however, to arrive through a
site-link transformation at an equivalent driven diffusive system in which
two types of oppositely charged particles diffuse asymmetrically and are
continuously created and annihilated in pairs.

In this work we have proceeded to a further investigation of the RSOS
version of the models first proposed in \cite{aehm}. Our study is based
on the mapping of the master equation governing the dynamics
of the associated reaction-diffusion process into an 
imaginary-time Schr\"{o}dinger equation, the Hamiltonian of which
is that of a spin $S=1$ non-Hermitian quantum chain \cite{chico}. This 
allows us to employ standard finite-size scaling (FSS) techniques to the
resulting stochastic process, in the same way as one is used to do with 
Hamiltonian theories or with the transfer matrices of classical spins systems
\cite{privman}. In this way we were able to analyze the time evolution
operator for chains of sizes up to 16 sites, calculating their spectra
and stationary states. The paper is organized as follows. In Sec.~II we 
present the basic formalism in which we work, comment on its physical contents
and derive the time evolution operator for the particular process we are
interested in. In Sec.~III we show and discuss the finite-size data we 
obtained for the dynamic critical exponent, the order parameter of the 
roughening transition, and the string order parameter. Finally, in Sec.~IV
we conclude and indicate some directions for further investigation.


\section{THE TIME EVOLUTION OPERATOR}

We are going to focus on the stochastic particle system associated with the
growth model, more specifically on its stochastic transition or intensity
matrix. There is quite a number of different ways of writing down the 
master equation in operator form, some more suited to the study of 
symmetries \cite{chico}, some others envisaging a perturbative 
approach \cite{peliti}. Here we give a brief derivation that parallels 
that of \cite{chico}.

Let us attach to each site $\ell$ of a one-dimensional lattice $\Lambda 
\subset {\Bbb Z}$ of volume $|\Lambda | = L$ a stochastic variable
$n_{\ell}$ taking values in the set of states $\omega = 
\{0,1,\ldots,N-1\}$. Denoting by $P({\bf n},t)$ the probability of realization
of a particular configuration ${\bf n} = (n_{1}, n_{2}, \ldots,
n_{L}) \in \Omega = \omega^{\Lambda}$ at instant $t$, we write the 
master equation as
\begin{equation}
\label{EM}
\frac{\partial P({\bf n},t)}{\partial t} = \sum_{\tilde{{\bf n}} \in \Omega}
\left[ \Gamma({\bf n},\tilde{{\bf n}})P(\tilde{{\bf n}},t)-
\Gamma(\tilde{{\bf n}},{\bf n})P({\bf n},t) \right],
\end{equation}
where $\Gamma(\tilde{{\bf n}}, {\bf n}) > 0$ is the rate for the 
transition ${\bf n} \to \tilde{{\bf n}}$. When only binary collisions 
intervene, we write $\Gamma(\tilde{{\bf n}}, {\bf n}) = \Gamma^{ab}_{cd}$ for
the elementary process $(a,b) \to (c,d)$, and the master equation reads
\begin{equation}
\label{ABCD}
\frac{\partial P({\bf n},t)}{\partial t} = \sum_{(\ell,m) \in \Lambda}
\left[ \sum_{a,b=0}^{N-1}\!\!^{\prime}\Gamma^{n_{\ell}+a,
n_{m}+b}_{n_{\ell},n_{m}}P(n_{1},\ldots,n_{\ell}+a,\ldots,n_{m}+b,\ldots,
n_{L},t)-\sum_{c,d=0}^{N-1}\!\!^{\prime}\Gamma^{n_{\ell},
n_{m}}_{n_{\ell}+c,n_{m}+d} P({\bf n},t) \right],
\end{equation}
\noindent
where the primes indicate that the free channel $(a,b) = (0,0) = (c,d)$ 
should not be considered in the summations and the additions in the 
indices of the rates and in the arguments of $P({\bf n},t)$ are all taken
modulo $N$. Periodic boundary conditions on $\Lambda$ will be understood in 
what follows.

We now introduce vector spaces in the description of Eq.~(\ref{ABCD}). To do 
this we turn $\omega = \{0,1,\ldots,N-1\}$ into $\omega = {\Bbb C}^{N}$
and ${\bf n}$ into $|{\bf n}\rangle = |n_{1},n_{2},\ldots,n_{L}\rangle \in 
\Omega = \omega^{\otimes\Lambda}$. Taking the orthonormal basis 
$\{|{\bf n}\rangle\}$ for $\Omega$ we write
\begin{equation}
\label{P}
|P(t)\rangle = \sum_{{\bf n} \in \Omega} P({\bf n},t)|{\bf n}\rangle
\end{equation}
for the generating vector of the probabilities $P({\bf n},t) =
\langle {\bf n} |P(t)\rangle$. We are in this way providing the space of
generating functions with a Hilbert space structure. Next we endow the
space of linear transformations of $\omega$ with the canonical basis of
matrices $E^{ab}$ with elements $(E^{ab})_{ij} = \delta_{ai}\delta_{bj}$,
$0 \leq a,b,i,j \leq N-1$, and introduce the operator $X$ such that $X_{ij} = 
\delta_{i+1,j}$, $X^{N} = {\bf 1}$. A little reflection and comparison show 
that within this settings we can write Eq.~(\ref{ABCD}) in the form
\begin{equation}
\label{ES}
\frac{d|P(t)\rangle}{dt} = -H|P(t)\rangle
\end{equation}
with
\begin{equation}
\label{OP}
H = \sum_{(\ell,m) \in \Lambda}\sum_{a,b=0}^{N-1}\!\!^{\prime}
\sum_{c,d=0}^{N-1}\!\!^{\prime} ({\bf 1}-X_{\ell}^{a-c}X_{m}^{b-d})
\Gamma^{ab}_{cd}\, E^{aa}_{\ell}E^{bb}_{m}.
\end{equation}
$H$ is but the infinitesimal generator of the Markov semigroup $U(t) =
\exp (-tH)$ of the continuous-time Markov chain defined by the set of rates
$\Gamma_{cd}^{ab}$. From Eq.~(\ref{OP}) we see that $H$ is a 
$N^{L}\times N^{L}$, usually nonsymmetric, and very sparse matrix. One says 
that Eq.~(\ref{ES}) is an imaginary-time Schr\"{o}dinger equation, the fact 
that $H$ has not the physical significance of energy and in general observes
$H^{\dagger} \neq H$ notwithstanding. It is however useful and intuitive to
think of $H$ as a Hamiltonian whose quantum fluctuations govern the time 
fluctuations of the classical system of particles.

The process we are interested in is a growth model defined as follows
\cite{aehm}. Let $h_{\ell} \in {\Bbb N}$ be the height of a surface
at site $\ell \in \Lambda$. The surface evolves by attempting, sequentially
and at randomly chosen sites, adsorption of an adatom $h_{\ell} \to 
h_{\ell}+1$ with probability $qdt$, and desorption of an adatom
$h_{\ell} \to \min \{h_{\ell -1},h_{\ell}\}$ or $h_{\ell} \to 
\min \{h_{\ell},h_{\ell +1}\}$ each with probability $\frac{1}{2}(1-q)dt$. 
We now impose the RSOS condition $|h_{\ell +1} - h_{\ell}| \leq 1, \forall
\ell \in \Lambda$, which suggests the use of link variables 
$c_{\ell} = h_{\ell +1}-h_{\ell} \in \{-1,0,1\}$. In order not to overload 
the notation with unnecessary pluses and minuses, let us take the values of
$c_{\ell}$ modulo 3, mapping $c_{\ell} = -1,0,1$ into $c_{\ell} = 2,0,1$, 
respectively. In this representation the growth process can be described by
the set of transition rates
\begin{equation}
\begin{array}{rclrcl}
\label{GAMMAS}
\Gamma^{10}_{01} & = & \frac{1}{2}(1-q), & \Gamma^{01}_{10} & = & q, \\
 & & & & &  \\
\Gamma^{20}_{02} & = & q, & \Gamma^{02}_{20} & = & \frac{1}{2}(1-q), \\
 & & & & &  \\
\Gamma^{12}_{00} & = & 1-q, & \Gamma^{21}_{00} & = & q, \\
& & & & &  \\
& & {\rm and} & \Gamma^{00}_{12} & = & q.
\end{array}
\end{equation}
According to Eq.~(\ref{OP}) we write for this process the time evolution
operator as
\begin{equation}
\label{MAT}
H=\sum_{\ell =1}^{L}H_{\ell,\ell+1}
\end{equation}
with the two-body stochastic transition matrix given explicitly by
\begin{equation}
H_{\ell,\ell+1} = \left( \begin{array}{ccccccccc}
q & \cdot & \cdot & \cdot & \cdot & -q & \cdot & -(1-q) & \cdot \\
\cdot & \frac{1}{2}(1-q) & \cdot & -q & \cdot & \cdot & \cdot & \cdot & \cdot \\
\cdot & \cdot & q & \cdot & \cdot & \cdot & -\frac{1}{2}(1-q) & \cdot & \cdot \\
\cdot & -\frac{1}{2}(1-q) & \cdot & q & \cdot & \cdot & \cdot & \cdot & \cdot \\
\cdot & \cdot & \cdot & \cdot & \cdot & \cdot & \cdot & \cdot & \cdot \\
\cdot & \cdot & \cdot & \cdot & \cdot & q & \cdot & \cdot & \cdot \\
\cdot & \cdot & -q & \cdot & \cdot & \cdot & \frac{1}{2}(1-q) & \cdot & \cdot \\
-q & \cdot & \cdot & \cdot & \cdot & \cdot & \cdot & 1-q & \cdot \\
\cdot & \cdot & \cdot & \cdot & \cdot & \cdot & \cdot & \cdot & \cdot 
\end{array} \right),
\end{equation}
\noindent
where we have ordered the two-site basis vectors antilexicographically, 
i.e., $|0,0\rangle \prec |1,0\rangle \prec \ldots \prec |1,2\rangle 
\prec |2,2\rangle$, and the dots indicate null entries.

Before proceeding to the next section, it is worth mentioning some properties
of $H$. As a stochastic transition matrix, conservation of probability flux
requires $\sum_{i}H_{ij} = 0$, which in turn constraint the diagonal elements 
to be given by $H_{jj} = -\sum_{i \neq j} H_{ij}$, compare with Eq.\ (\ref{EM}).
The first of these conditions imply the existence of a trivial left eigenstate
with zero eigenvalue,
\begin{equation}
\label{OMEGA}
|\Omega\rangle = \sum_{{\bf n} \in \Omega} |{\bf n}\rangle, \qquad 
\langle \Omega |H = 0,
\end{equation}
with $\langle \Omega | \Omega \rangle = N^{L}$ the cardinality of
the state space. This special vector plays a role in the determination 
of expectation values, for one can write the average of an observable 
$A({\bf n})$ with respect to the probabilities $P({\bf n},t)$ with the aid 
of $|\Omega\rangle$ as
\begin{equation}
\label{EXPECT}
\langle A \rangle (t) = \sum_{{\bf n} \in \Omega}A({\bf n})P({\bf n},t) =
\langle \Omega |A|P(t)\rangle.
\end{equation}
We expect physical observables of the classical system of particles to be 
diagonal in the natural basis $\{|{\bf n}\rangle\}$, once they all have to
commute. $H$ is obviously not diagonal in this basis, it is not an observable
of the system. Eq.~(\ref{EXPECT}) summarizes an important difference between
quantum physics and the kind of classical physics we have here: it is that 
expectation values are linear in $|P(t)\rangle$, not bilinear, the elements
of $|P(t)\rangle$ being probabilities themselves, not probability amplitudes.

Besides these general properties, our $H$ in Eq.~(\ref{MAT}) is additionally
translation invariant, due to the homogeneous rates in a ring geometry, and
possesses a U(1) symmetry labeled by the total charge $Q=Q^{(+)}-Q^{(-)}$, 
which is conserved along the process. These symmetries allow us to 
block-diagonalize $H$ and to write it as the direct sum
\begin{equation}
\label{SECTS}
H = \sum_{Q=-L}^{L}\sum_{k=0}^{L-1}H^{Q}_{k} ,
\end{equation}
where $Q$ and $k$ label the U(1) and momentum eigensectors, respectively.
Each U(1) sector represents a closed class of the stochastic process, 
the corresponding block matrix $H^{Q}$ being itself a stochastic transition
matrix governing the dynamics within the given sector. We say that our process
is decomposable, nonergodic, and that the U(1) label classifies its $2L+1$
irreducible, closed classes. The momentum label $k$, however, is introduced
here solely in order to take advantage, numerically, of the further reduction
of order $1/L$ on the sizes of the blocks $H^{Q}$ it furnishes, the physically
relevant momentum sector being the one with $k=0$, since the zero 
eigenvalue together with the completely flat surface of the model is in this 
sector. One can look for a relation $E(k) \propto k^{\theta}$ for the real part
of the eigenspectrum, but we do not expect to extract useful information,
e.g., about $\theta$, from this relation for small lattice sizes. Moreover,
the interpretation of such a relation for the low lying excitations of the
process as a dispersion relation for quasiparticles of a free field theory
would be somewhat cavalier; see \cite{henkel,dennijs}.


\section{FINITE-SIZE SCALING}

According to Eqs.~(\ref{GAMMAS}), as $q$ increases creation of $+ -$ as well 
as annihilation of $- +$ pairs increases while the remaining processes induce 
segregation of particles. This corresponds to an increase in adsorption and in
the growth rate of islands, leading to rougher configurations. As $q$ lowers, 
increased annihilation of $+ -$ pairs together with more symmetric diffusion 
of 0's flatten the surface; see Sec.~III~C\@. Around the critical point 
$q=q_{c}$ separating these two phases the correlation lengths of the infinite
system behave like
\begin{equation}
\label{CORREL}
\xi_{\|} \,\propto\, \xi_{\perp}^{\theta} \,\propto\, |q-q_{c}|^{-\nu_{\|}}
\,\propto\, |q-q_{c}|^{-\nu_{\perp}\theta}.
\end{equation}
Here $\xi_{\|}$ is the correlation length in the time direction
while $\xi_{\perp}$ is the one in the spatial direction, with
$\nu_{\|}$ and $\nu_{\perp}$ the corresponding critical exponents and
$\theta = \nu_{\|}/\nu_{\perp}$ the dynamic critical exponent.

For finite systems of size $L$, according to the usual FSS assumptions
\cite{privman,kinzel,deseze}, we expect the scaling
\begin{equation}
\label{CSIL}
\xi_{\|, L} \propto L^{\theta_{L}},
\end{equation}
to hold when $q=q_{c,L}$, the finite version of the critical point $q_{c}$,
with $\theta_{L}$ the finite version of $\theta$. On general grounds one
expects $\lim_{L \to \infty} q_{c,L} = q_{c}$ and $\lim_{L \to \infty}
\theta_{L} = \theta$. Therefore from Eqs.~(\ref{CORREL}) and (\ref{CSIL})
we obtain the relations
\begin{equation}
\label{FSS}
      \frac{\ln \left[ \xi_{\|,L}(q_{c,L})/\xi_{\|,L-1}(q_{c,L}) \right]}
           {\ln (L/L-1)} =
      \frac{\ln \left[ \xi_{\|,L+1}(q_{c,L})/\xi_{\|,L}(q_{c,L}) \right]}
           {\ln(L+1/L)}  = \theta_{L},
\end{equation}
which through the comparison of three successive system sizes furnishes 
simultaneously $q_{c,L}$ and $\theta_{L}$. Of course $\xi_{\|, L}^{-1}=
{\rm Re}\{\Delta E_{L}\}$, with $\Delta E_{L} = E_{L}^{(1)}$ the first gap 
of $H$, since for stochastic transition matrices $E_{L}^{(0)}=0$ by 
construction.

\subsection{Dynamic critical exponent}

We have calculated the gaps $\Delta E_{L}$ in the $Q=0$, $k=0$ sector through 
the use of the Arnoldi algorithm \cite{arnoldi,golub}. This is a Krylov 
subspace projection technique that effects the reduction of a general 
nonsymmetric matrix to upper Hessenberg form, the eigenpairs of which converge
variationally to that of the original matrix with the algebraically larger 
part of the spectrum converging first. In order to save memory we have used a 
restarted version of the algorithm, in which we fix the dimension of the Krylov
subspace and use some of the approximate eigenvectors obtained in one iteration
as the starting vectors for the next iteration, until convergence is obtained 
to the desired accuracy. In this way we were able to handle matrices of orders
up to $324\,862$ with up to $\sim \! 8 \times 10^{6}$ nonzero entries keeping 
the Krylov subspace always with fewer than 64 vectors.

The results we have obtained are summarized in Tabs.\ \ref{tab1} and 
\ref{tab2}. The extrapolated values in the last line of these tables were 
obtained through the Bulirsch-Stoer extrapolation scheme \cite{bst}, with
$\omega_{\rm BST}$ the free parameter of the algorithm chosen over a 
certain range so as to optimize the converge of the finite-size data.

\begin{figure}
\begin{center}
\mbox{\epsfig{figure=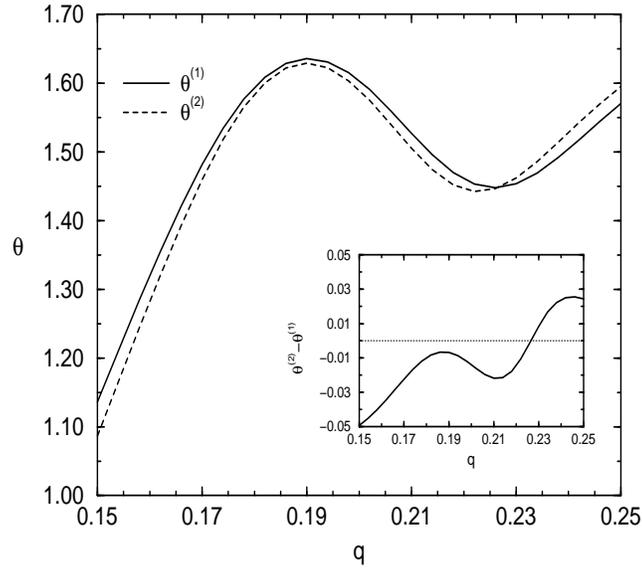,height=8.6cm,width=8.6cm,angle=-90}}
\end{center}
\caption{Variation of the exponent $\theta$ with $q$. The values of 
$\theta^{(1)}$ and $\theta^{(2)}$ were obtained from the first and second
expressions in Eqs.\ (\ref{FSS}), respectively, using the triplet of
lengths $L-1,L,L+1 = 13,14,15$. The inset shows that the difference
between their values reaches a minimum around $q=0.188$ and vanishes
around $q=0.226$. The finite-size sequences obtained from both the minima
and the crossings converge to well defined limit values, see Tables \ref{tab1}
and \ref{tab2} and the text.}
\label{fig1}
\end{figure}

In applying Eqs.~(\ref{FSS}) we found two consistent, converging sets of
data, the first one, shown in Table~\ref{tab1} and marked with a prime,
realizing the first equality in Eqs.~(\ref{FSS}) only approximately, and the
second one, that in Table~\ref{tab2} and marked with two primes, realizing it
exactly. This behavior is illustrated in Fig.~\ref{fig1}.

The first set of data exhibited a rather smooth convergence in both the values
of $q_{c,L}$ and $\theta_{L}$, while the second set behaved more irregularly.
The values in Table~\ref{tab1} indicate a second order transition taking place
around $q_{c}' \simeq 0.1875$ with a dynamic exponent of $\theta' \simeq 
1.585$. This value of $\theta'$ is compatible with the exponent of the directed
percolation process, for which the most accurate value to date, obtained 
by Monte Carlo simulations, seems to be $\theta_{\rm DP} = 1.580\,75(3)$ 
\cite{jensen}; it should be mentioned that in an evaluation
of $\theta_{\rm DP}$ more closely related to ours it has been found a value
of $\theta_{\rm DP} = 1.588(1)$ \cite{avraham}, and that it is not clear why
such discrepancies arise in the value of $\theta_{\rm DP}$ calculated by
different methods.

The interpretation of the data in Table~\ref{tab2} is touchier. We can see that
the values of $q_{c,L}''$ converge at a reasonable rate to the extrapolated
limit $q_{c}'' \simeq 0.1932$, which is different from the previously
found $q_{c}' \simeq 0.1875$. We believe however that if we have had access
to larger lattice sizes, we would have observed $q_{c}'' \to q_{c}'$, since
the values of $q_{c}'$ vary less.  We thus trust the value of 
$q_{c}' = 0.1875(1)$ as our best estimate for the critical point.
The situation with the critical exponent $\theta_{L}''$
is different: it seems to be converging to a completely different value than 
$\theta_{\rm DP}$.  In fact, this behavior was to be expected, for it
has been found \cite{aehm} that in the rough phase $q > q_{c}$ the 
exponents of the process are those of the Kardar-Parisi-Zhang
universality class \cite{kpz}, in particular $\theta_{\rm KPZ}=\frac{3}{2}$.
As can be seen from Table~\ref{tab2}, the first few values of 
$\theta_{L}''$ show a monotonic increasing behavior up to 
$\theta_{9}'' \simeq 1.4649$, but then the sequence begins to
decrease to reach the bottom value of $\theta_{15}'' \simeq 1.4426$.
A partial extrapolation of the first four points, $6 \leq L \leq 9$, gives
$\theta'' =1.47$, while a partial extrapolation of the rest of the points,
$10 \leq L \leq 15$, gives $\theta'' =1.42$. This absence of monotonicity 
of the finite-size data for $\theta_{L}''$ is quite unusual, and we have not
yet a clear clue to this behavior. Even so, we can take the set of values of
$\theta_{L}''$ in Table~\ref{tab2} as indicating the presence of a critical 
region for $q > q_{c}$ with an exponent $\theta'' \neq \theta_{\rm DP}$, 
possibly with $\theta'' =\theta_{\rm KPZ}$.

Concerning the exponent $\nu_{\perp}$, we found it not possible to apply 
the standard approach \cite{kinzel,deseze} to obtain its value because the 
derivatives of ${\rm Re}\{ \Delta E_{L}\}$ with respect to $q$ evaluated at 
the points $q_{c,L}$ change sign for some pairs of lattice sizes, thus 
preventing us from taking logarithms. The finite-size sequences obtained with
the absolute values of these derivatives as well as with those obtained with
the derivatives of the absolute values of the gaps also failed to converge, 
so that we were not able to obtain an estimate of $\nu_{\perp}$ from our 
diagonalizations.

\subsection{Spontaneous symmetry breaking}

Given that the surface suffers a transition from a flat phase to a rough 
phase, it is natural to think of an order parameter which measures this
transition. Moreover, it is interesting to have an order parameter taking into
account the symmetries of the process, which are besides the translational
and U(1) symmetries, a Z$_{\infty}$ symmetry related to the fact that the
microscopic dynamics is invariant under an arbitrary integer shift
$h_{\ell} \to h_{\ell}+n$ in the heights. A proper order parameter is given 
by \cite{aehm} 
\begin{equation}
\label{ORDER}
M_{L}=\frac{1}{L}\sum_{\ell =1}^{L}(-1)^{h_{\ell}} ,
\end{equation}
which is fast in the sense of not being conserved by
the dynamics. The choice of such an order parameter anticipates the 
interpretation of the roughening transition (actually, of the flattening 
transition) as the result of a spontaneous break of the Z$_{\infty}$ symmetry,
for while in the rough phase all heights are exploited evenly, in the flat 
phase the system spontaneously selects one fiducial level around which the 
heights fluctuate. One then expects $M_{L}$ to be finite in the flat phase
while vanishing in the rough phase due to canceling fluctuations.

\begin{figure}
\begin{center}
\mbox{\epsfig{figure=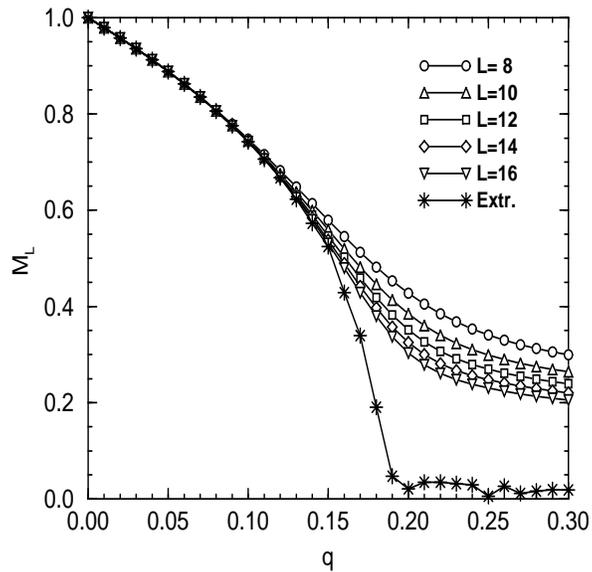,height=8.6cm,width=8.6cm,angle=-90}}
\end{center}
\caption{Order parameter $M_{L}$ for even $L$ between $L=8$ and $L=16$ 
together with the extrapolated curve.}
\label{fig2}
\end{figure}

We have calculated $M_{L}$ for even $L$ between $L=8$ and $L=16$;
our finite-size data together with the extrapolated values appear in 
Fig.~\ref{fig2}. From that figure we clearly see the transition taking place
around $q=0.190$, although the precise determination of the critical point is
not possible from this figure. We have not found the signature of two different
$q_{c}$'s in our data for $M = \lim_{L \to \infty}M_{L}$, which we regard as 
an indication that $q_{c}''$ should indeed tend to $q_{c}'$ as $L \to \infty$.

\begin{figure}
\begin{center}
\mbox{\epsfig{figure=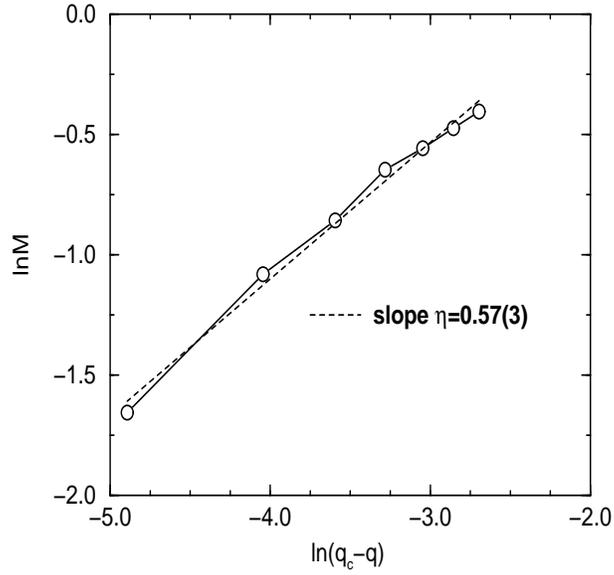,height=8.6cm,width=8.6cm,angle=-90}}
\end{center}
\caption{Plot of $\ln M \times \ln (q_{c}-q)$. The LR slope gives
the critical exponent $\eta = 0.57 \pm 0.03$.}
\label{fig3}
\end{figure}

The order parameter $M$ vanishes around $q \lesssim q_{c}$ as 
$M \propto (q_{c}-q)^{\eta}$. The plot of $\ln M \times \ln (q_{c}-q)$ for
the points $0.12 \leq q \leq 0.18$ of the extrapolated curve in 
Fig.\ \ref{fig2} together with a linear regression fit appear in 
Fig.\ \ref{fig3}. We found an exponent $\eta = 0.57 \pm 0.03$, which compares 
well with previous results in the literature: in the first of the papers by 
Alon {\it et al.\/} \cite{aehm}, $\eta$ has been evaluated as $0.55 \pm 0.05$,
while in a recent simulation of a model of yeastlike growth of fungi colonies 
with parallel dynamics it has been found that $\eta \simeq 0.50$ \cite{lopez}.
Also in a certain line in the phase diagram of a one-dimensional 
next-nearest-neighbor asymmetric exclusion process closely related to these 
growth models it has been found that $\eta = 0.54 \pm 0.04$ \cite{nnnasep}.
In the second of the papers by Alon {\it et al.\/} \cite{aehm}, however, the 
more accurate value $\eta = 0.66 \pm 0.06$ has been published, pushing the 
estimate to a somewhat higher value. Recent preliminary Monte Carlo 
simulations of ours, on the other hand, suggest a typical Ginzburg-Landau 
scenario for the symmetry break in these models, which would then predict an
$\eta = \frac{1}{2}$. We believe that more extensive simulations can settle
this point, and work is being done in this direction.

\subsection{String order parameter}

In order to better understand the nature of the roughening transition,
let us look at some typical microscopic configurations of the model.
The roughest possible surface is given in the links representation by 
$|+ + \cdots + - - \cdots - \rangle$, the state of maximal height in the 
$Q=0$ sector of the dynamics. In spin language, this state corresponds 
to two domains separated by two antiferromagnetic (AF) kinks. The second 
roughest possible surface configurations are given by those with a pair
of 0's, e.g., $|+ \cdots + 0 \, 0 - \cdots - \rangle$ or
$|0 + \cdots + - \cdots - 0 \rangle$. From this example and the rates 
in Eqs.\ (\ref{GAMMAS}) it becomes clear that the flattening process
is induced by AF kink annihilation, while diffusion of 0's introduce surface
shape fluctuations. It is important to notice that absence of desorption
from the middle of smooth terraces enforces a certain AF order among the 
particles, for pairs are created only as $+ -$ pairs, never as $- +$ pairs, and 
since there are no $+ - \rightleftharpoons - +$ reactions (which would violate
the RSOS condition), we see that this order persists as long as pairs survive
annihilation. In the rough, high $q$ phase the $+$ particle will preferably
move leftwards, while the $-$ particle will prefer to move to the right,
eventually leaving a pair of 0's in between which then generates another
$+ -$ pair, thus leading to rough configurations like, e.g.,
$|\cdots + + 0 \cdots - 0 - \cdots \rangle$. In the flat, low $q$ phase
diffusion of particles becomes more symmetric and segregation of particles
less efficient, and we thus expect to observe a more uniform distribution
of $+ -$ pairs along the lattice than in the high $q$ phase. The completely
flat surface without any AF kink is only attained at $q=0$.

In the context of two-dimensional RSOS crystal growth models and the Haldane
conjecture \cite{haldane}, it has been predicted \cite{string} and subsequently
extensively verified \cite{arovas,tasaki,kennedy,hatsugai,alcaraz,white} 
that a particular type of long-range order exists in the spin $S=1$ 
antiferromagnetic isotropic Heisenberg (AFH) chain. In the ground state of the
AFH chain, this order may be viewed as made up of (not necessarily closely)
bound $+ -$ dipoles interspersed among the 0's, forming what has become known
as an AF spin fluid. The order parameter that identifies this type of order is
the so-called string order parameter 
\cite{string},
\begin{equation}
\label{SOP}
O^{z}_{\pi}(\ell) = \left\langle 
S^{z}_{1} \exp \left[ i\pi \sum_{n=1}^{\ell} S^{z}_{n} \right] S^{z}_{\ell}
                    \right\rangle ,
\end{equation}
where the brackets indicate expectation value in the ground state. 
In the gapped, Haldane phase of the AFH chain as well as in the disordered
flat phase of the models in \cite{string} one has 
$\lim_{\ell \to \infty}O^{z}_{\pi}(\ell) \neq 0$.

\begin{figure}
\begin{center}
\mbox{\epsfig{figure=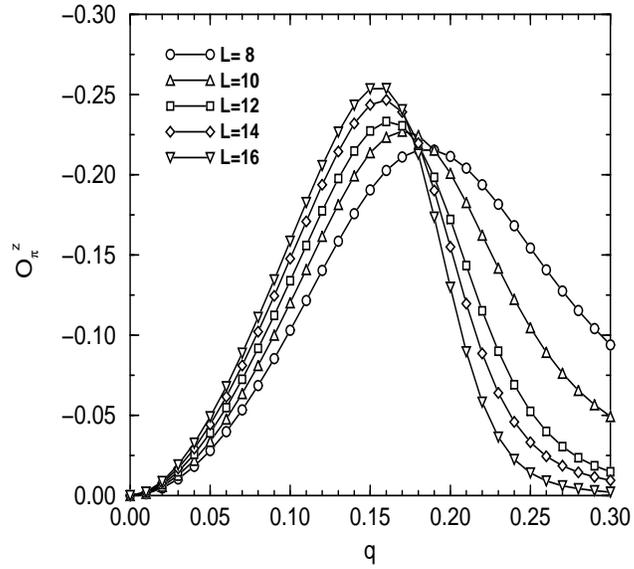,height=8.6cm,width=8.6cm,angle=-90}}
\end{center}
\caption{String order parameter $O^{z}_{\pi}(L/2+1)$ for even $L$ 
between $L=8$ and $L=16$.}
\label{fig4}
\end{figure}

From what we said above it is clear that the string order is just the kind
of order we expect to observe in the flat phase of our model. We have thus
calculated the steady state expectation value of the string order parameter
(\ref{SOP}) with $S^{z}_{n} = c_{n}$ and $\ell = L/2+1$, the maximum
distance in a ring geometry, and the results appear in Fig.~\ref{fig4}.
Our extrapolations did not perform well for this set of data, and are not
showed in this figure. The general trend, however, is quite clear: above
the critical point, $q > q_{c}$, $O^{z}_{\pi}(\ell)$ strongly tends to zero,
while for $q \leq q_{c}$ we have $\lim_{\ell \to \infty}O^{z}_{\pi}(\ell) 
\neq 0$. The point $q=0$ is special, for at $q=0$ the completely flat surface
becomes an absorbing state and $O^{z}_{\pi}(\ell)=0$ exactly at this point.
As $q$ grows from zero, the AF spin fluid begins to form and 
$O^{z}_{\pi}(\ell)$ grows accordingly, until at $q=q_{c}$ the asymmetry in 
the diffusion rates for the particles disrupts this AF spin fluid structure,
ordered domains begin to prevail and the string order vanishes. We then see
that the string order parameter clearly reveals the mechanism of the
roughening transition as the unbounding of the fluid antiferromagnetic
pairs in favor of the formation of ordered domains.


\section{SUMMARY AND CONCLUSIONS}

In summary, we carried out a finite-size scaling study of the
roughening transition in a class of one-dimensional RSOS models which also 
presents spontaneous symmetry breaking. We found that at the critical point 
$q_{c} = 0.1875(1)$ the transition occurs with a
dynamic exponent compatible with that of the directed site percolation
process, for which we have the estimate $\theta = 1.585(1)$, and that above
$q_{c}$ there is a critical rough phase most probably with KPZ exponents.
Unfortunately, we were unable to calculate a second exponent from our
diagonalizations. This might be due to the nonhermiticity of the operator
$H$, which might have caused an unusual nonmonotonicity in the values
of the gaps with the parameter $q$, thus preventing us from obtaining 
$\nu_{\perp}$. This lack of monotonicity has already been reported in the
literature \cite{arndt,dmrg}, where it has also been noticed the slow 
convergence of the finite-size data towards the infinite volume limit.

The order parameter $M$ was found to vanish like $M \propto (q_{c}-q)^{\eta}$
for $q \lesssim q_{c}$ with an exponent $\eta = 0.57 \pm 0.03$, in agreement
with previously found values of $\eta$ \cite{aehm,lopez,nnnasep}. The 
calculation of the string order parameter revealed that the flat, gapped phase
of the model is a disordered phase analogous to a Haldane phase, with the 
stationary state presenting an antiferromagnetic spin-fluid structure of kinks,
although dominated by the completely flat surface with no such a structure. 
The roughening transition may thus be understood in the links representation 
as the unbounding of the fluid antiferromagnetic pairs in favor of the 
formation of ordered domains, which then begin to blend together providing 
the surface with a finite growth velocity.

It is possible to push further the investigation of this class of 
models in one definite way. The idea is to allow for an explicit break
of the symmetry in the set of rates Eq.~(\ref{GAMMAS}) by the following
artifact \cite{ritt}. In the particle scenario, we double the number of
sites, introducing between two successive links $c_{\ell},c_{\ell+1}$ a
noninteracting flag variable $m_{\ell +\frac{1}{2}}$ taking two possible
values, call them~$+$ and~$-$. This variable will mimic the pseudospin
$(-1)^{h_{\ell}}$. We then allow the rates of our modified model to be 
parametrized by, besides $q$, a chiral symmetry breaking field 
$u \in [-1,1]$ such that now the rates depend on the quantities
\[
p=(1-u)q \qquad {\rm and} \qquad \tilde{p}=(1+u)q
\]
according to $\Gamma^{a+b}_{c-d}=\Gamma^{ab}_{cd}(p)$ and 
$\Gamma^{a-b}_{c+d}=\Gamma^{ab}_{cd}(\tilde{p})$, the roles of $p$ and
$\tilde{p}$ interchanged whenever $\Gamma^{ba}_{dc} \neq 0$ for a given
$\Gamma^{ab}_{cd} \neq 0$. For example, 
$\Gamma^{1+0}_{0-1} = \frac{1}{2}(1-p)$ and
$\Gamma^{1-0}_{0+1} = \frac{1}{2}(1-\tilde{p})$, but
$\Gamma^{0+1}_{1-0} = \tilde{p}$ and $\Gamma^{0-1}_{1+0} = p$.
The choice of which combination of values of $m$'s in the new rates
will pick a $p$ or a $\tilde{p}$ with respect to the original rates 
is immaterial, for letting $u \to -u$ exchange their roles. For this
process one may look at the order parameter
\[
M_{L}(u)=\frac{1}{L}\sum_{\ell =1}^{L}m_{\ell +\frac{1}{2}}
\]
to see whether one finds a spontaneously symmetry-broken phase.
It may happens that for some values of the field $u$ one gets spinodal
points, and that these points are associated with unusual dynamic
exponents, e.g.~$\theta =1$, once they have already appeared in one
dimensional driven diffusive systems \cite{arndt}. This problem is 
currently under investigation.


\section*{ACKNOWLEDGMENTS}

The author would like to acknowledge Professor Vladimir Rittenberg for having
called his attention to this subject and Professor Francisco C. Alcaraz for
helpful advice and many suggestions that greatly improved the final form of 
the manuscript. This work was supported by Funda\c{c}\~{a}o de Amparo \`{a}
Pesquisa do Estado de S\~{a}o Paulo (FAPESP), Brazil.



\begin{table}
\centering
\caption{Finite-size data for the region where a minimum is
observed. The numbers between parentheses represent the estimated
errors in the last digit of the data.}
\label{tab1}
\begin{tabular}{d|d|d}
$L-1,L,L+1$ & $q_{c,L}'$ & $\theta_{L}'$     \\ \hline
5,6,7    & 0.190$\,$461(1) & 1.794$\,$411(1) \\
6,7,8    & 0.190$\,$676(1) & 1.749$\,$008(1) \\
7,8,9    & 0.189$\,$988(1) & 1.718$\,$350(1) \\
8,9,10   & 0.189$\,$294(1) & 1.695$\,$320(1) \\
9,10,11  & 0.188$\,$753(1) & 1.677$\,$170(1) \\
10,11,12 & 0.188$\,$355(1) & 1.662$\,$589(1) \\
11,12,13 & 0.188$\,$076(1) & 1.650$\,$766(1) \\
12,13,14 & 0.187$\,$885(1) & 1.641$\,$112(1) \\
13,14,15 & 0.187$\,$761(1) & 1.633$\,$190(2) \\
14,15,16 & 0.187$\,$683(1) & 1.626$\,$655(2) \\ \hline
Extrapolated & 0.1875(1)   & 1.585(1)        \\
{[}$\omega_{\rm BST}${]} & [7.329] & [2.543] \\
\end{tabular}
\end{table}

\begin{table}
\centering
\caption{Finite-size data for the region where crossing occurs.
The data without an associated error are believed to be correct 
to the figures shown.}
\label{tab2}
\begin{tabular}{d|d|d}
$L-1,L,L+1$ & $q_{c,L}''$ & $\theta_{L}''$   \\ \hline
5,6,7    & 0.295$\,$749 & 1.445$\,$967       \\
6,7,8    & 0.275$\,$660 & 1.458$\,$235       \\
7,8,9    & 0.261$\,$781 & 1.463$\,$535       \\
8,9,10   & 0.251$\,$648 & 1.464$\,$937       \\
9,10,11  & 0.244$\,$053 & 1.463$\,$830       \\
10,11,12 & 0.238$\,$186 & 1.461$\,$059       \\
11,12,13 & 0.233$\,$511 & 1.457$\,$203       \\
12,13,14 & 0.229$\,$689 & 1.452$\,$672       \\
13,14,15 & 0.226$\,$497 & 1.447$\,$753       \\
14,15,16 & 0.223$\,$784 & 1.442$\,$647       \\ \hline
Extrapolated & 0.1932(1) & {\rm see text}       \\
{[}$\omega_{\rm BST}${]} & [1.815] & {\rm see text} \\ 
\end{tabular}
\end{table}


\end{document}